# Cosmic Strings and Anisotropic Universe

K.L. Mahanta[1] and S.K. Tripathy[2]

[1] Department of Mathematics, C. V. Raman College of Engineering,
Mahura, Janla, Bhubaneswar, Odisha, India.
e-mail: kamal2_m@yahoo.com

[2] Department of Physics, Indira Gandhi Institute of Technology, Sarang, Dhenkanal, Odisha, India, 759146.
e-mail: tripathy_sunil@rediffmail.com

**Abstract**

Plane symmetric bulk viscous string cosmological models with strange quark matter are investigated. We have incorporated bulk viscous pressure to study its affect on the properties of the model. Assuming an anisotropic relationship among the metric potentials, we have tried to put some constraints on the anisotropic parameter.

## 1. Introduction

In astrophysics and cosmology the matter distribution can be suitably described by a relativistic perfect fluid in equilibrium due to the large scale distribution of galaxies in our universe. However there are several processes like interaction between matter and radiation [1], quarks and gluons [2] and different components of dark matter [3] where such a perfect fluid model is not appropriate. These processes involve dissipation which can be described in terms of a bulk viscous fluid. Further a decay of massive superstring modes into massless modes [4], gravitational string production [5, 6] and particle creation effects in the grand unification era [7] are the processes expected to give rise to the viscous effect. Bulk viscosity is another irreversible phenomenon which contributes to entropy production in the universe. It is believed that dissipative processes in the early stages of the evolution of the universe may well account for the high degree of isotropy and the ratio of the number of photons to baryons. Another advantage of the introduction of bulk viscosity is that we may get singularity free solutions. Bali and Dave [8] studied the behavior of Bianchi type III string cosmological model in presence and absence of bulk viscous fluid in general relativity. Wang [9] assumed bulk viscosity as a power function of scalar of expansion and constructed Kantowski-Sachs bulk viscous string cosmological model.

Bali and Pradhan [10], Tripathy et al. [11, 12], Rao et al. [12], Kandalkar et al. [13] investigated different cosmological models in the presence of bulk viscosity and cosmic string.

Over a period of time, there have many attempts for the modifications of EGTR to facilitate the unification of gravitation and many other effects in the universe. Barber [15] proposed two continuous self-creation theories by modifying the Brans-Dicke [16] theory and general theory of relativity. In the modified theories of Barber, the universe is created out of self-contained gravitational and matter fields. In the first theory the solution of the one body problem yields unsatisfactory results. This theory cannot be derived using action principle and more importantly, in this theory the equivalence principle is violated [17]. However, the second theory is a modification of EGTR incorporating a variable Newtonian gravitational constant G. In this theory the gravitational coupling of the Einstein's field equations is considered to be a variable scalar on the space-time manifold. In addition, this scalar does not gravitate directly but simply divide the matter tensor and acts as a reciprocal gravitational constant. This theory is capable of verification or falsification which can be done by observing the behavior of degenerate matter and photons. In view of consistency of the second theory in this paper we intend to construct and study cosmological model in this theory.

String theory was developed to describe events at the early stages of the evolution of the universe. In gauge theories with spontaneously broken symmetries strings arise as a random network of stable line like topological defects during the phase transitions which occur when the temperature decreases below the critical temperature at the early stages of the evolution of the universe [18]. Massive closed loops of string serve as seeds for the formation of large structures like galaxies and cluster of galaxies. While matter is accreted onto loops, they oscillate violently and lose their energy by gravitational radiation and therefore they shrink and disappear. If string exists, they can produce a number of characteristic observational effects detectable with existing astronomical instruments. Kaiser and Stebbins [19] pointed out that such strings would produce density fluctuations on very large scales and may be responsible for the formation of large scale structures. The gravitational effects of cosmic strings are responsible for creation of galaxies and clusters [20, 21]. Letelier [22] investigated Bianchi type I and Kantowski-Sachs string cosmological models which evolve from a pure massive string dominated era to a particle dominated era, with or without remnant of strings. Venkateswarlu et al. [23] studied Bianchi type III, VIII and IX string cosmological models in Second Self-creation Theory (SST) of gravitation. Mohanty and Mahanta [24], Rao et al. [25] and Rao and Vinutha [26] are some of the authors who have investigated various string cosmological models in Barber's SST.

In the early phase of the evolution of the universe it is believed that quark-gluon plasma existed when the temperature was ~200MeV. The two ways of formation of quark matter are the quark- hadrons phase transition and conversion of neutron stars into strange ones at ultra high densities [27-29]. The equation of state based on the phenomenological bag model of quark matter where quark confinement is described by an energy term proportional to the volume is used to model the strange quark matter. In this model quarks are considered as degenerate Fermi gas, which exist only in a region of space endowed with a vacuum energy density $B_C$. Moreover in this model the quark matter is composed of light u, d quarks, massive s quarks and electrons. In our present study, we have considered a simplified version of this model with massless and non-interacting quarks.

In presence of quark matter in the universe, the total matter energy density is considered to consists of quark energy density $\rho_q$ and the vaccum energy density $B_C$,

$$\rho_m = \rho_q + B_c \tag{1}$$

and the total pressure is

$$P_m = P_q - B_c \tag{2}$$

where, the quark pressure is expressed as

$$P_q = \frac{\rho_q}{3}. \tag{3}$$

Yilmaz [30] and Adhav et al. [31] constructed Kaluza-Klein cosmological models for string and domain wall with quark matter in EGTR. Mahanta et al. [32] studied Bianchi type III string cosmological model with quark matter in Barber's SST of gravitation. Recently Rao and Neelima [33] investigated strange quark matter attached to string cloud in Barber's SST and EGTR in the axially symmetric space time. Sahoo and Mishra studied string cloud and domain wall with quark matter for a plane symmetric metric in biometric theory [34].

Considering the importance of bulk viscosity in cosmological models with respect to the accelerated expansion phase of the universe, in this work we construct bulk viscous string cloud cosmological model with strange quark matter attached to one dimensional cosmic strings in Barber's SST and EGTR. Without considering any functional form for bulk viscosity, we have tried to get viable cosmological model.

The organization of the paper is as follows: in sect-2, assuming a negative constant deceleration parameter, we present the general field equations for a quark filled universe where the contribution to pressure comes only from bulk viscosity and discuss their exact solutions. Sect-3 contains a discussion on the Barber's scalar field. The physical properties of the models both in the presence and absence of bulk viscosity are discussed in sect-4. At the end, summary and conclusion of the works are presented in sect-5.

## 2. Anisotropic plane symmetric model

The universe is observed to be flat and homogeneous in the large scale and can be well described by FRW metric. However, anisotropic Bianchi type models are being investigated in recent times and are interesting in the sense that they are more general than FRW model. Bianchi type models are based on exact solutions of Einstein's general relativity and are homogeneous but not necessarily having spatial isotropy. Anisotropic models can account for certain large scale anomalies of the isotropic standard cosmologies such as (i) detection of large-scale velocity flows significantly larger than those predicted in standard cosmology, (ii) a statistically significant alignment and planarity of Cosmic microwave background(CMB) quadrupole and octupole modes and (iii) the observed large scale alignment in quasar polarization [35, 36]. Precise measurements of Wilkinson Microwave Anisotropy Probe (WMAP) show the unusual alignment and planarity of the quadrupole and octupole which suggests that one direction of the universe may have expanded differently from the other two transverse directions [37, 38] which signals non trivial topology of the large scale geometry of the universe [39]. It has been shown in a recent work that, if the large scale spatial geometry of our universe is plane symmetric with an eccentricity at decoupling of order of $10^{-2}$, the quadrupole amplitude can be reduced drastically without affecting higher multipoles of the angular power spectrum of the temperature anisotropy [40]. For a planar symmetry, the universe looks the same from all the points but the points all have preferred axis. However, it may be noted here that, there still persists uncertainty on these large angle anisotropies and they remain as open problems. Keeping this in view, we consider the plane symmetric space-time metric in the form

$$ds^2 = dt^2 - A^2(dx^2 + dy^2) - B^2 dz^2 \qquad (4)$$

where A and B are the directional scale factors considered as functions of cosmic time t only. The metric corresponds to considering $xy$-plane as the symmetry plane. The eccentricity of such a universe is given by $e = \sqrt{1 - \frac{B^2}{A^2}}$. The field equations in Barber's SST are

$$G_i^j \equiv R_i^j - \frac{1}{2}g_i^j R = -\frac{8\pi}{\varphi} T_i^j \qquad (5)$$

and

$$\Box\varphi = \frac{8}{3}\pi\eta T \qquad (6)$$

where η is the coupling constant to be evaluated from experiment and φ is the Barber's scalar. Here φ is a function of t only. In the limit η → 0, the theory approaches Einstein's theory in every respect.

The energy momentum tensor for a cloud of string dust with bulk viscosity compatible with the metric (4) is given by

$$T_{ij} = \rho u_i u_j - \rho_s x_i x_j - \xi u^l_{;l}(u_i u_j - g_{ij}) \qquad (7)$$

where $\rho$ is the total rest energy density for the cloud of strings with particles attached to them, $\rho_s$ is the string tension density, $\xi$ is the bulk coefficient of viscosity, $u^i$ the four velocity of the particles, $x^i$ the unit space like vector representing the direction of the string. Here $u^i$ and $x^i$ satisfy the orthogonal relationship

$$u^i u_i = -x^i x_i = 1 \tag{8}$$

and

$$u^i x_i = 0 \tag{9}$$

The total rest energy density is the sum of the energy density due to particles $\rho_p$ and the string tension,

$$\rho = \rho_p + \rho_s. \tag{10}$$

We assume that the universe consists of massless non-interacting quarks of all possible flavours which contribute to the energy density due to particles, $\rho_p = \rho_q + B_c$ and hence,

$$\rho = \rho_q + \rho_s + B_c \tag{11}$$

The field equations in the self creation theory can now be explicitly written as

$$\left(\frac{A'}{A}\right)^2 + 2\frac{A'B'}{AB} = \frac{8\pi}{\varphi}\rho \tag{12}$$

$$\frac{A''}{A} + \frac{A'B'}{AB} + \frac{B''}{B} = \frac{8\pi}{\varphi}\xi\theta \tag{13}$$

$$2\frac{A''}{A} + \left(\frac{A'}{A}\right)^2 = \frac{8\pi}{\varphi}(\rho_s + \xi\theta) \tag{14}$$

and

$$\varphi'' + \theta\varphi' = \frac{8\pi}{3}\eta(\rho + \rho_s + 3\xi\theta) \tag{15}$$

where a dash over the field variables represents ordinary time derivative and $\theta$ is the expansion scalar given by

$$\theta = 2\frac{A'}{A} + \frac{B'}{B}. \tag{16}$$

Considering the highly non linear in nature of the field equations (12) – (16), we require some additional relations to get viable cosmological solutions. Following Bali [41], we assume that expansion ($\theta$) is proportional to the components of shear tensor ($\sigma$) which leads to the anisotropic relation between the metric potentials

$$B = A^n \tag{17}$$

where $n(> 0)$ is an arbitrary constant that takes care of the anisotropic nature of the model. If $n = 1$, the model reduces to be isotropic and anisotropic otherwise. However, we are interested in an anisotropic model and we consider any other value of the anisotropic parameter else than unity. The mean Hubble expansion rates can be expressed in terms of the expansion rates along the longitudinal and transverse directions as, $H = \frac{R'}{R} = \frac{1}{3}\left(2\frac{A'}{A} + \frac{B'}{B}\right)$ where $R = (A^2 B)^{1/3}$ is the average scale factor. The deceleration parameter is expressed as

$$q = -1 - \frac{H'}{H^2}. \tag{18}$$

A positive value of $q$ envisages a decelerating universe where as a negative $q$ leads to an accelerated expansion of the universe. Recent observations of distant type Ia Supernovae predict that in the present epoch the universe is not only expanding but the expansion is accelerating [42-44]. Such an observational fact suggests a negative deceleration parameter which is predicted to be $q_0 = -0.81 \pm 0.14$ in the present time [45]. In attempts to predict of the accelerated expansion of the universe, the sign and nature of deceleration parameter is considered in different manner in different theoretical works. The time variation of the deceleration parameter is still under intense debate, even though in certain models, a time varying $q$ leads to transient phenomena from early deceleration to late time acceleration [46-50]. It is to worth mention here that the special law of variation of Hubble's parameter as proposed by Berman [51, 52] yields a constant deceleration parameter. In view of this, in the present work, we consider a negative value for the constant deceleration parameter to mimic the accelerated expansion phase of the universe.

With a constant deceleration parameter $q = q_0$, eq. (18) yields

$$R = [(1 + q_0)t + b]^{\frac{1}{1+q_0}} \tag{19}$$

where $b$ is a constant of integration. It is evident from (19) that an expanding universe favours $1 + q_0 > 0$. In other words, an accelerated expansion phase can be obtained from the present model if deceleration parameter lies in the range $0 > q_0 > -1$. For $q_0 = -1$, the mean Hubble parameter becomes constant and we get a de Sitter universe with an exponential scale factor. It is straight forward to get the metric potentials from the radius scale factor in (19) using (17),

$$A = [(1+q_0)t + b]^{\frac{3}{(1+q_0)(n+2)}} \tag{20}$$

and

$$B = [(1+q_0)t + b]^{\frac{3n}{(1+q_0)(n+2)}}. \tag{21}$$

The line element for the plane symmetric metric (4) for a constant deceleration parameter can now be written in terms $q_0, b$ and the anisotropic exponent $n$ as

$$ds^2 = dt^2 - [(1+q_0)t + b]^{\frac{6}{(1+q_0)(n+2)}}(dx^2 + dy^2) - [(1+q_0)t + b]^{\frac{6n}{(1+q_0)(n+2)}} dz^2. \tag{22}$$

### 3. Barber's Scalar field

The mean Hubble parameter $H$ and the scalar expansion $\theta$ for this model can be expressed as

$$H = \frac{1}{(1+q_0)t+b}, \tag{23}$$

$$\theta = \frac{3}{(1+q_0)t+b}. \tag{24}$$

In the present epoch, $t = t_0$, $H = H_0$ and $\theta = \theta_0$. Hence in terms of values in the present epoch, we can express

$$H = \frac{H_0}{(1+q_0)(t-t_0)H_0 + 1}, \tag{25}$$

$$\theta = \frac{\theta_0}{(1+q_0)(t-t_0)H_0 + 1}, \tag{26}$$

and the radius scale factor

$$R = R_0[(1 + q_0)(t - t_0)H_0 + 1]^{1/(1+q_0)}, \tag{27}$$

where $R_0$ is the present value of the radius scale factor.

From the field eqs. (12) - (15) we find

$$[(1 + q_0)t + b]^2 \varphi'' + \beta(1 + q_0)[(1 + q_0)t + b]\varphi' - \eta\alpha(1 + q_0)^2 \varphi = 0 \tag{28}$$

where, $\beta = \frac{3}{1+q_0}$ and $\alpha = \frac{6(n^2+2n+3)}{(1+q_0)^2(n+2)^2} - \frac{2}{1+q_0}$. Barber's scalar field can now be expressed as

$$\varphi = c_1[(1 + q_0)t + b]^{m_1} + c_2[(1 + q_0)t + b]^{m_2} \tag{29}$$

where $m_1 = \frac{1}{2}\left((1 - \beta) + \sqrt{(\beta - 1)^2 + 4\eta\alpha}\right)$ and $m_2 = \frac{1}{2}\left((1 - \beta) - \sqrt{(\beta - 1)^2 + 4\eta\alpha}\right)$, $c_1$ and $c_2$ are the integration constants.

Here we observe that when the coupling constant $\eta \to 0$, one mode of $\varphi$ i.e. $[(1 + q_0)t + b]^{m_2} \nrightarrow const.$ provided $\beta \neq 1$. However for $\beta = 1$, the scalar field comes out to be a constant quantity. Since $n$ is positive and $q_0$ is considered to be negative, $\beta$ should always be greater than 3. In fact, we consider here the deceleration parameter to be equal to the prediction from observations of type 1a Supernovae at high redshift i.e. $q_0 = -0.8$ [45], and hence we get, $\beta = 15$. In view of this, the mode of $\varphi$ corresponding to the exponent $m_2$ is not acceptable. Thus we have

$$\varphi = c_1[(1 + q_0)t + b]^{m_1}. \tag{30}$$

In terms of Hubble parameter at the present epoch, the scalar field can be expressed as

$$\varphi = c_1 \left[\frac{(1+q_0)(t-t_0)H_0+1}{H_0}\right]^{m_1}, \tag{30}$$

which clearly shows that in the present epoch the scalar field becomes $\varphi_0 = \frac{c_1}{H_0^{m_1}}$ and we have

$$\varphi = \varphi_0[(1 + q_0)(t - t_0)H_0 + 1]^{m_1}. \tag{31}$$

The Barber scalar $\varphi \to \infty$ as $t \to \frac{(1+q_0)t_0-1}{(1+q_0)H_0}$ and $\varphi \to 0$ as $t \to \infty$ if $m_1 < 0$. For $m_1 < 0$, with the increase in cosmic time, the Barber scalar field decreases from its initial value [1 −

$(1 + q_0)H_0 t_0]^{m_1}$. In this case of the exponent $m_1$, the SST is significant at the early stages of the evolution of the universe and the theory leads to EGTR at a later epoch for large cosmic time. However, for $m_1 > 0$, the Barber scalar $\varphi$ increases with the growth of cosmic time.

## 4. Physical properties of the model

The physical and geometrical parameters for the plane symmetric metric (22) with cosmic strings composed of strange quark matter are obtained as

String tension density:
$$\rho = \frac{9(1+2n)H_0^{2-m_1}}{8\pi(n+2)^2[(1+q_0)(t-t_0)H_0+1]^{2-m_1}} \tag{32}$$

String tension density:
$$\rho_s = \left[\frac{18-3(n+2)(1+q_0)-9n-9n^2+3n(n+2)(1+q_0)}{8\pi(n+2)^2[(1+q_0)(t-t_0)H_0+1]^{2-m_1}}\right]H_0^{2-m_1} \tag{33}$$

Quark energy density:
$$\rho_q = \rho - B_c = \frac{9(1+2n)H_0^{2-m_1}}{8\pi(n+2)^2[(1+q_0)(t-t_0)H_0+1]^{2-m_1}} - B_c \tag{34}$$

Quark pressure:
$$P_q = \frac{\rho_q}{3} = \frac{9(1+2n)H_0^{2-m_1}}{8\pi(n+2)^2[(1+q_0)(t-t_0)H_0+1]^{2-m_1}} - \frac{B_c}{3} \tag{35}$$

Bulk viscous coefficient:
$$\xi = \frac{[3(n^2+n+1)-(n^2+3n+2)(1+q_0)]}{8\pi(n+2)[(1+q_0)(t-t_0)H_0+1]^{1-m_1}}H_0^{1-m_1} \tag{36}$$

Shear scalar:
$$\sigma^2 = \left(\frac{1-n}{n+2}\right)^2 \frac{3H_0^2}{[(1+q_0)(t-t_0)H_0+1]^2} \tag{37}$$

These properties, in general, depend on the anisotropic exponent n, deceleration parameter $q_0$ and the coupling constant $\eta$. Obviously there occurs a big-bang at $t = t_0 - \frac{5}{H_0}$. If we correlate this to be at the beginning of the universe i.e. at $t = 0$, the age parameter of the universe can be well determined through the deceleration parameter as $H_0 t_0 = 5$. If $H_0 t_0 > 5$, it can be inferred that at the beginning of cosmic time there is some residual volume which decrease with time and vanishes at $t = t_0 - \frac{5}{H_0}$ and then again expansion occurs. If $H_0 t_0 < 5$, the big crunch may be believed to have occurred before. Since $n \geq 0, \rho \geq 0$, in other words, the anisotropic model we have considered satisfies the energy condition. The scalar of expansion is infinitely large at $t = t_0 - \frac{5}{H_0}$ and it tends to vanish at large cosmic time. The tensor of rotation $\omega_{ij} = u_{i,j} - u_{j,i}$ is zero for the model. Hence the model represents a shearing and non-rotating universe with a possible big crunch at some initial epoch. The string energy density, bulk viscous coefficient, string tension density, quark energy density and quark pressure become infinite at $t = t_0 - \frac{5}{H_0}$ for $m_1 < 2$ and approach to zero when $t \to \infty$. For $m_1 > 2$, these properties of the universe increase with the growth of cosmic time. Since $\lim_{t \to \infty} \frac{\sigma^2}{\theta^2} \neq 0$, the model does not approach isotropy for large $t$.

Moreover, when the coupling constant $\eta \to 0$, this theory leads to EGTR and the model represents a bulk viscous string cosmological model with strange quark matter attached in Einstein's general relativity. In this case, $m_1$ vanishes and the scalar field becomes a constant. Like that of the properties in the SST model, all of the properties in EGTR will have a point singularity at $t = t_0 - \frac{5}{H_0}$. These properties evolve through the cosmic time and decrease from large value in initial phase to null values at infinitely large cosmic time. We can have an evolving relationship between the energy density and the bulk viscosity of the model as

$$\rho \propto \frac{\xi}{(t-t_0)H_0+1} \tag{38}$$

which implies that, the energy density of the universe decreases faster than the bulk viscosity with the evolution of the universe.

In the absence of bulk viscosity, the field equation (13) reduces to

$$-\frac{\dot{H}}{H^2} = \frac{3(n^2+n+1)}{n^2+3n+2}. \tag{39}$$

The right hand side of this eqn.(39) will give us a positive quantity for any value of the anisotropic exponent $n$. The deceleration parameter, in the absence of bulk viscosity can be calculated from (39) to be $q = \frac{2(n^2-1)}{n^2+3n+2}$ which can assume negative value if the anisotropic exponent assumes values less than unity. In other words, in the absence of bulk viscosity, in the plane symmetric self creation model, we achieve accelerated expansion of the universe only if $n$ lies in the range $0 < n < 1$. If we accept that the deceleration parameter should be as that of the observationally predicted value, then we can put a tight constraint on the value of the anisotropic exponent as $n = \frac{1}{14}$. However, it should be noted here that, the exact determination of the deceleration parameter requires the observation of objects with redshift greater than one which is a formidable task in the present time and therefore, current observational results on deceleration parameter are not reliable [53]. In such a case of anisotropic parameter, the eccentricity of the universe increases very fast and approaches unity in a short time. This is not acceptable which implies a necessity of inclusion of bulk viscosity in the model. In the absence of bulk viscosity, the Barber's scalar field is expressed as $\varphi = c_1 t^{m_1}$, where $m_1 = \frac{1}{2}\left((1-\alpha) + \sqrt{(\alpha-1)^2 + 4\eta\beta}\right)$, $\alpha = \frac{(n+2)(n+1)}{n^2+n+1}$ and $\beta = \frac{2(n+1)^3}{3(n^2+n+1)^2}$. The rest energy density and string tension density in the absence of bulk viscosity can be evaluated as

$$\rho = \frac{c_1(n+1)^2(1+2n)}{8\pi(n^2+n+1)^2 t^{2-m_1}}, \tag{40}$$

$$\rho_s = \frac{c_1(n+1)^2}{8\pi(n^2+n+1)^2 t^{2-m_1}}, \tag{41}$$

The behavior of the properties of the model without bulk viscous pressure is almost the same as that of with bulk viscosity. In the absence of bulk viscosity, big-bang singularity occurs at the beginning. An accelerated expansion phase of universe with all plausible parameters of the universe is not possible in the absence of bulk viscosity.

## 5. Summary and Conclusion

In the present work, we have investigated plane symmetric and homogeneous cosmological model in Barber's second self creation theory. The matter content of the universe is assumed to have a cloud of one dimensional cosmic strings with particles attached to them. The particles are considered to consist of strange quark matters. As dissipative phenomena in the anisotropic cosmic fluid we have incorporated bulk viscosity. The pressure in the fluid comes from the bulk viscosity part only. In most of the earlier investigations, bulk viscosity is assumed to be a power function of energy density or a power function of the scalar of expansion. In this work we have obtained the models in Barber's SST and EGTR without considering such functional forms for bulk viscosity. In the present work, we have restricted ourselves to a constant negative deceleration parameter as predicted from observations. We found that in the discussed model, $\xi \propto [(1+q)(t-t_0)H_0 + 1]\rho$, whereas $\xi \propto \theta^{1-m_1}$ in SST. In the early stages of the evolution of the universe bulk viscosity exists and has a greater role in the production of energy density of the universe. In the absence of bulk viscosity, it is not possible to get viable accelerating cosmological model. The string tension density and quark matter density are found to depend upon the anisotropic parameter as well as the deceleration parameter. Obviously, the Barber's coupling constant has bearing in the properties of the model. In the absence of bulk viscosity the big bang singularity occurs right at the beginning of the cosmic time, where as, in the presence of there occurs a possibility of big crunch at some initial epoch, not exactly at the beginning. It can be emphasized that the inclusion of bulk viscosity enriches the output and leads to interesting results.